\documentclass[twocolumn,prl,reprint,letterpaper]{revtex4-1}
\usepackage{graphicx}% Include figure files
\usepackage{dcolumn}% Align table columns on decimal point
\usepackage{bm}% bold math
\usepackage{amsmath}
\usepackage{amssymb}
\usepackage{amsfonts}
\usepackage{color}
\usepackage{psfrag}
\usepackage{tensor}
\newcommand{\ud}{\mathrm{d}}

\begin{document}

%\preprint{APS/123-QED}

\title{Asymptotic Green's function for the stochastic reproduction of competing variants via Fisher's angular transformation}%
%\title{Asymptotic diallelic Green's function for population genetics under selection and neutrality via Fisher's angular transformation}% Force line breaks with \\

\author{Bhavin S. Khatri}

\affiliation{%
The Francis Crick Institute, Mill Hill Laboratory,\\
The Ridgeway, London, NW7 1AA, U.K.
}%

\date{\today}% It is always \today, today,
%Abstract character limit of 600 characters

%\maketitle
\begin{abstract}

The Wright-Fisher Fokker-Planck equation describes the stochastic dynamics of self-reproducing, competing variants at fixed population size. We use Fisher's angular transformation, which defines a natural length for this stochastic process, to remove the co-ordinate dependence of it's diffusive dynamics, resulting in simple Brownian motion in an unstable potential, driving variants to extinction or fixation. This insight allows calculation of very accurate asymptotic formula for the Green's function under neutrality and selection, using a novel heuristic Gaussian approximation.

\end{abstract}
\maketitle

Understanding the interplay between stochastic and deterministic forces in systems with different reproducing variants is a theme that arises, and has importance, in many different scientific fields \cite{Blythe2007} including language evolution \cite{Baxter2006}, protein evolution \cite{Kimura1984,Akashi2012}, the evolution of biodiversity \cite{HubbellBook2001,McKane2000,Volkov2003} and population genetics \cite{Crow1970,Wright1945}. This article is concerned with a fundamental question in population genetics, given the possibility of only two reproducing variants, how does the probability distribution of gene frequency $x(t)$ change over time, given it is known at a prior time point $x_0=x(0)$, subject to small number fluctuations (genetic drift) and selection (competition), though the analogous question may be posed in any of these fields. We address this question in the context of the Wright-Fisher model, which is the canonical model of stochastic dynamics incorporating both these features.

The diffusion approximation \cite{Kimura1964}, of the Wright-Fisher model describes the stochastic dynamics of gene frequency $x$ ($=n/N$, where $n$ is the number of copies of the mutant allele and $N$ the total population):

\begin{align}\label{Eq:ForwardFokkerPlanck}
\frac{\partial G(x,x_0;t)}{\partial t}=-\frac{\partial }{\partial x}&\left(sx(1-x)G(x,x_0;t)\right) \\
&+\frac{1}{2N}\frac{\partial^2 }{\partial x^2}\left(x(1-x)G(x,x_0;t)\right),\nonumber
\end{align}
where $G(x,x_0;t)$ is the transition probability density, or Green's function, of gene frequency given an initial condition $G(x,x_0;0)=\delta(x-x_0)$, and $s$ is the selection coefficient. This equation is derived in the large $N$ limit from a Master equation of discrete populations of each variant at a fixed $N$ \cite{Kimura1964}.

This Fokker-Planck equation has been studied extensively. In particular, Kimura calculated a series solution for the neutral equation in terms of Gegenbauer polynomials \cite{Kimura1954,Kimura1964}, which was later extended to the multi-allele case by Baxter, et al \cite{Baxter2007}. For the case of selection Kimura also calculate a series solution, however, the eigenvalues could not be represented in closed-form in terms of the population size $N$ and selection coefficient $s$ \cite{Kimura1964}. More recently, a number of methods have been developed to calculate the Green's function under selection, including a numerical matrix approach \cite{Song2012} and perturbation theory based on a path-integral formulation of the Wright-Fisher process \cite{Schraiber2014}. However, for many practical applications, such as virus evolution, where population sizes are large and generation times short, these approaches are not very practical as a large number of terms is required for convergence at short times. The solution of Voronka and Keller \cite{Voronka1975}, which uses an asymptotic ray approximation of the Green's function, is valid at short times and for models of selection, neutrality and mutation, but their approach is not very intuitive and unwieldy requiring switching between different solutions in a time-dependent manner. We present a simple short-time asymptotic calculation of the Green's function in closed form for both neutrality and selection, which has intuitive appeal as it exploits Fisher's angular transformation, which as we show is the natural co-ordinate for Wright-Fisher stochastic dynamics \cite{Antonelli1977}.

%We present a simple calculation of the Green's function in closed form for both neutrality and selection, which has intuitive appeal as it exploits Fisher's angular transformation, which gives a natural co-ordinate for the stochastic dynamics, where diffusion is co-ordinate independent and any co-ordinate dependence in diffusion is manifested through a non-zero convective/drift term. Fisher's angular transformation is a specific example of more general approach in higher dimensions that develops a relationship between stochastic dynamics and Riemannian geometry, where the effective metric tensor is the inverse of the covariance matrix of the stochastic process and the natural or ``stochastic'' distance between two points is defined by the metric \cite{Antonelli1977}.

%For neutrality and weak selection we make a harmonic approximation of the effective potential that arises after Fisher's angular transformation. For strong selection, we develop a novel heuristic method of finding Green's function solutions; this a priori assumes a Gaussian form, but with a time dependent variance calculated from the local curvature of the effective potential. The local curvature at a given time is determined from a solution of the mean from the deterministic differential equation in angular space.

%As far as the authors are aware there are no closed-form Green's function solutions, under neutrality or selection, that are accurate in the short time regime.

%\section*{Neutral Green's function}

\begin{figure}[htb!]
\begin{center}
{\rotatebox{0}{{\includegraphics[width=0.5\textwidth]{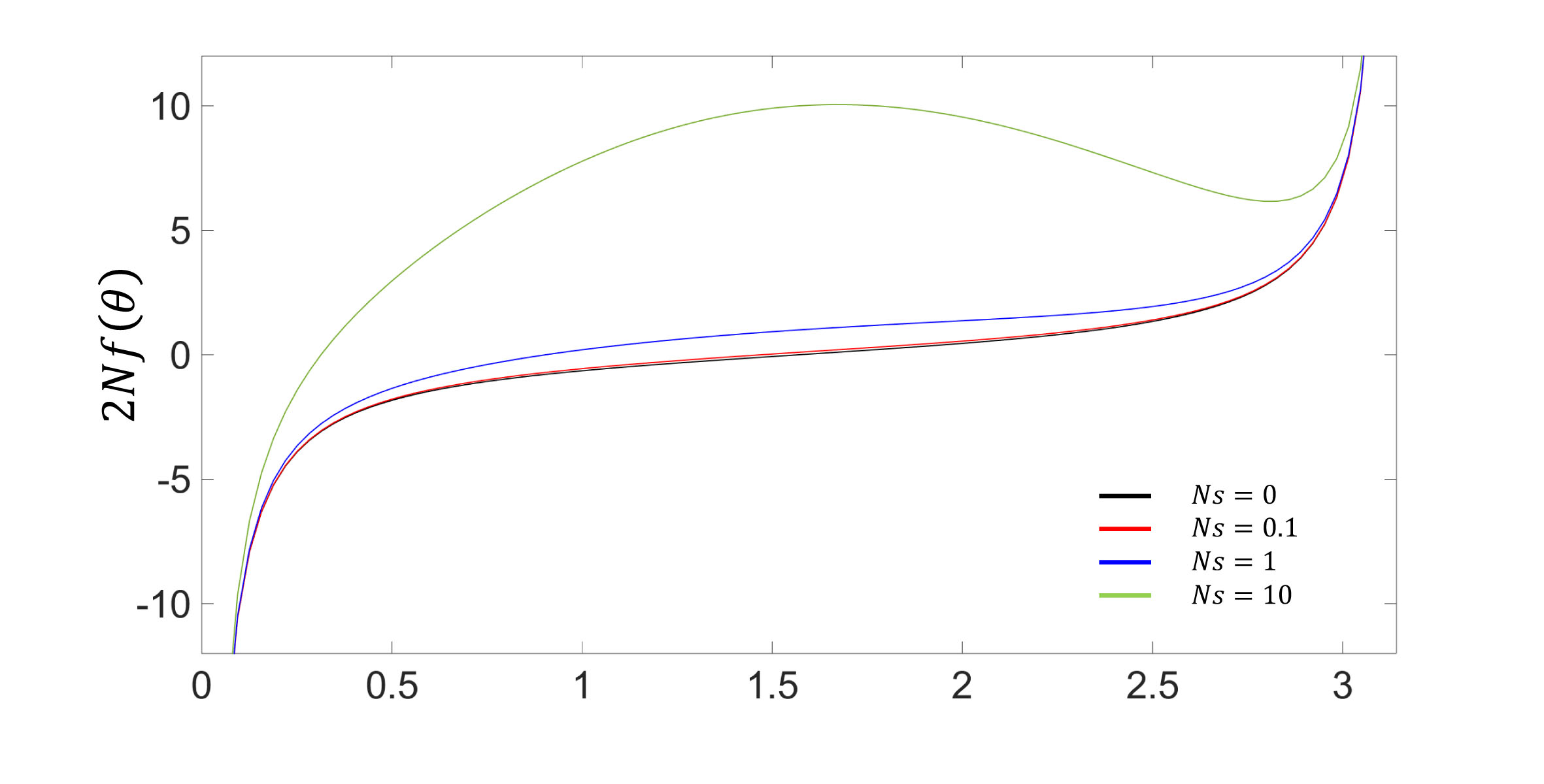}}}}
\caption{Effective deterministic force in angular domain for Wright-Fisher process, where $2NF(\theta)=-\cot\theta+Ns\sin\theta$.\label{Fig:Neutral+SelectiveForce}}
\end{center}
\end{figure}

Fokker-Planck equations with co-ordinate dependent diffusion constants such as Eqn.\ref{Eq:ForwardFokkerPlanck} have the property that space is explored at different rates dependent on the position in the domain; using this intuition Antonelli et al \cite{Antonelli1977}, suggested the natural definition of length for a stochastic process be related to the differential $\ud\theta^2=\sum_{ij}g_{ij}\ud x^i\ud x^j$, where $g_{ij}$ is a metric tensor and taken to be the inverse of the covariance matrix $g^{ij}$. In one-dimension, this is simply $\ud\theta^2=\ud x^2/D(x)$, which represents the (differential) mean square distance traversed in equal times. For the diffusion constant of random drift, $D(x)=x(1-x)$, the stochastic distance is simply

\begin{equation}\label{Eq:Stochastic_distance_Drift}
\theta=\int^x\frac{\ud x'}{\sqrt{x'(1-x')}}=\cos^{-1}(1-2x).
\end{equation}
This is Fisher's angular transformation \cite{Fisher1922,Fisher1930a}. Applying this transformation to the neutral Fokker-Planck equation ($s=0$), we get

%\begin{equation}\label{Eq:FisherAngularTransf}
%\theta=\cos^{-1}(1-2x).
%\end{equation}
%Applying this transformation to the neutral Fokker-Planck equation ($M=0$), gives \cite{Fisher1922,Fisher1930a}

\begin{equation}\label{Eq:PDE_Theta}
\frac{\partial p(\theta,t)}{\partial t}=\frac{1}{2N}\left(\frac{\partial^2 p(\theta,t)}{\partial \theta^2} +\frac{\partial }{\partial \theta}\left(\cot\theta p(\theta,t)\right)\right),
\end{equation}
or the equivalent stochastic differential equation \cite{GardinerBook, vanKampen}
\begin{equation}\label{Eq:SDE_Theta}
\frac{\ud\theta}{\ud t}=-\frac{1}{2N}\cot(\theta)+\eta(t),
\end{equation}
where $\langle\eta(t)\rangle=0$ and $\langle\eta(t)\eta(t')\rangle=\delta(t-t')/N$. We see the result of this transformation is a co-ordinate independent diffusion constant in $\theta-$space, but now with an effective deterministic force $f(\theta)=\cot(\theta)/2N$. This arises due to the co-ordinate dependent diffusion constant $x(1-x)/N$ in the $x-$domain and drives the system towards regions of decreasing diffusion constant. It is also exactly the spurious drift term that arises in transforming between Ito and Stratonovich descriptions of stochastic dynamics \cite{GardinerBook,vanKampen}. Note that in the Langevin representation, Eqn.\ref{Eq:ForwardFokkerPlanck} would have a multiplicative noise term, which is transformed to additive noise in $\theta-$space in Eqn.\ref{Eq:SDE_Theta}. Examining the force in Fig.\ref{Fig:Neutral+SelectiveForce}, we see that it is unstable, on average driving a mutant alelle to extinction if $\theta(0)<\pi/2$ and fixation if $\theta(0)>\pi/2$, with a fixed point at $\theta=\pi/2$. To calculate an approximate solution of the Green's function, we make a Taylor expansion of the force about the fixed point $\theta=\pi/2$ to linear order to give a linear stochastic differential equation, $\dot{\theta}=\frac{1}{2N}\left(\theta-\frac{\pi}{2}\right)+\eta(t)$. As we will see this approximation works well even for initial frequencies of order $10\%$ ($x(0)=0.1$), due to the non-linearity of the angular transformation, which has the property of compressing the central range in $x-$space about $x=1/2$ to a smaller central region in $\theta-$ space about $\theta=\pi/2$ (for example, $x=0.1\rightarrow\theta=0.64$ and $x=0.9\rightarrow\theta=2.5$). Equivalently, this is an harmonic approximation of the effective potential function in $\theta-$space, where $\dot{\theta}=\partial_\theta U(\theta) +\eta(t)$ and $U(\theta)=\frac{1}{2N}\ln\sin\theta$. As the resulting SDE is linear the solution is straightforwardly computed as $\theta(t)=\frac{\pi}{2}+\left(\theta_0-\frac{\pi}{2}\right)e^{t/2N}+\int_0^t\ud t'\eta(t')e^{\frac{t-t'}{2N}},$ where $\theta_0=\theta(0)$. As the solution is a sum of Gaussian random variables $\eta$ the Green's function for $\theta$ will be Gaussian with mean, $\langle\theta(t)\rangle=\frac{\pi}{2}+\left(\theta_0-\frac{\pi}{2}\right)e^{t/2N}$, since $\langle\eta(t)\rangle=0$ and variance, $\langle\langle\theta^2(t)\rangle\rangle=e^{t/N}-1$, where the van Kampen notation has been used, $\langle\langle \theta^2\rangle\rangle=\langle\theta^2\rangle-\langle\theta\rangle^2$. Note that the variance diverges for $t\gg N$ and the mean divergences for $t\gg 2N$, to $-\infty$ when $\theta_0<\pi/2$ and to $+\infty$ for $\theta_0>\pi/2$ and is fixed for all time at $\langle\theta\rangle=\pi/2$, if $\theta_0=\pi/2$ the fixed point of the deterministic dynamics. The Green's function in $\theta-$space is then

%\begin{equation}\label{Eq:Theta_Harmonic_Soln}
%\theta(t)=\frac{\pi}{2}+\left(\theta_0-\frac{\pi}{2}\right)e^{t/2N}+\int_0^t\ud t'\eta(t')e^{\frac{t-t'}{2N}},
%\end{equation}
%where $\theta_0=\theta(0)$. As the solution is a sum of Gaussian random variables $\eta$ the Green's function for $\theta$ will be Gaussian with mean, $\langle\theta(t)\rangle=\frac{\pi}{2}+\left(\theta_0-\frac{\pi}{2}\right)e^{t/2N}$, since $\langle\eta(t)\rangle=0$ and variance, $\langle\langle\theta^2(t)\rangle\rangle=e^{t/N}-1$, where the van Kampen notation has been used, $\langle\langle \theta^2\rangle\rangle=\langle\theta^2\rangle-\langle\theta\rangle^2$. Note that the variance diverges for $t\gg N$ and the mean divergences for $t\gg 2N$, to $-\infty$ when $\theta_0<\pi/2$ and to $+\infty$ for $\theta_0>\pi/2$ and is fixed for all time at $\langle\theta\rangle=\pi/2$, if $\theta_0=\pi/2$ the fixed point of the deterministic dynamics. The Green's function in $\theta-$space is then

\begin{align}\label{Eq:GreensFuncTheta_Harmonic_Soln}
G_\theta(\theta,\theta_0;t)&=\frac{1}{\sqrt{2\pi(e^{t/N}-1)}}\nonumber\\
&\times\exp{-\frac{\left(\theta-\theta_0e^{t/2N}-\frac{\pi}{2}(1-e^{t/2N})\right)^2}{2(e^{t/N}-1)}}.
\end{align}
Note the similarity of form to the Green's function of an overdamped harmonic oscillator, but with the difference that, as discussed, here the mean and variance diverge \cite{DoiEdwards}. This solution does not obey the boundary conditions at $\theta=0$ and $\theta=\pi$, which are required to be absorbing and specifically to go linearly to zero at these points; this is in order for the solution in $x-$space to be finite at the boundaries, as required due to the singularity of the diffusion constant at $x=0$ and $x=1$ \cite{Baxter2007}. The method of images cannot be used in this case as the required image has it's forces reversed and so does not obey the original Fokker-Planck equation. However, as we argue in the discussion, for many applications, including virus evolution, the short time behaviour ($t\ll N$) is most relevant. Transforming back to $x-$space, we have,

\begin{widetext}
\begin{align}\label{Eq:GreensFunc_x_Harmonic_Soln}
G_x(x,x_0;t)&=\left|\frac{\ud\theta}{\ud x}\right|G_\theta(\theta(x),\theta_0(x_0);t)\nonumber\\
&=\frac{1}{\sqrt{2\pi x(1-x)(e^{t/N}-1)}}\exp{-\frac{\left(\cos^{-1}(1-2x)-\cos^{-1}(1-2x_0)e^{t/2N}-\frac{\pi}{2}(1-e^{t/2N})\right)^2}{2\left(e^{t/N}-1\right)}},
\end{align}
\end{widetext}
where the Jacobian is $\left|\frac{\ud\theta}{\ud x}\right|=2/\sin\theta=1/\sqrt{x(1-x)}$.

\begin{figure}[htb!]
\begin{center}
{\rotatebox{0}{{\includegraphics[width=0.5\textwidth]{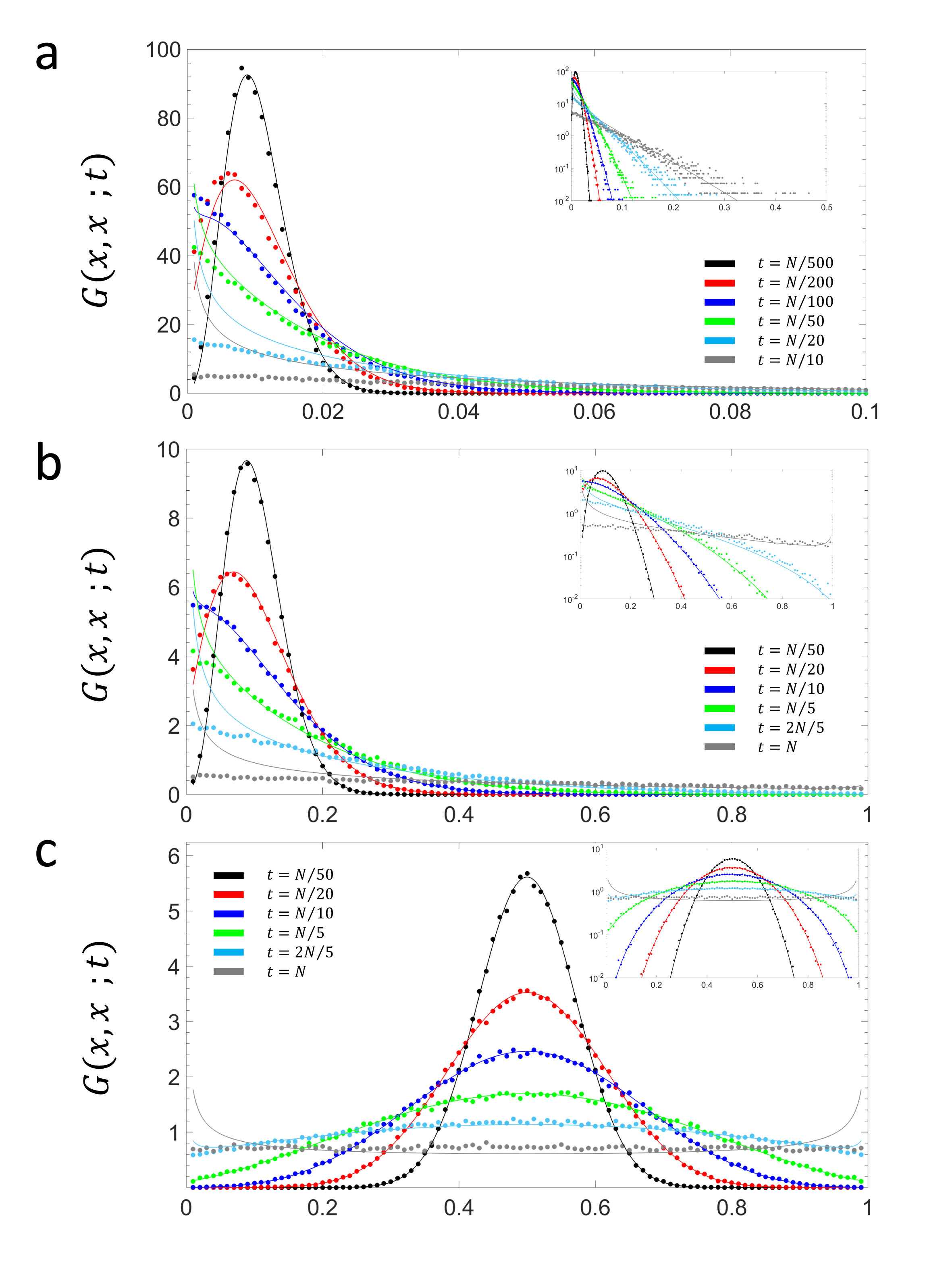}}}}
\caption{Comparison of approximate calculation of neutral Greens function (solid lines -- Eqn.\ref{Eq:GreensFunc_x_Harmonic_Soln}) and numerical integration of stochastic differential equation that arises from diffusion approximation (solid circles). a) initial frequency $x_0=0.01$, b) $x_0=0.1$, c) $x_0=0.5$.\label{Fig:NeutralGreensFunc}}
\end{center}
\end{figure}

The results are plotted in Fig.\ref{Fig:NeutralGreensFunc}, at various times and initial conditions, as solid lines and compared against numerical integration of the exact neutral Wright-Fisher stochastic differential equation (Eqn.\ref{Eq:SDE_Theta}). We see that, in general, the approximation works very well for short times $t\ll N$ and when the initial frequency $x_0$ is not too small. More precisely we would expect the approximation to be good for sufficiently short times compared to the expected time to fixation, which is $\langle t^*\rangle=-2N(x_0\ln(x_0)+(1-x_0)\ln(1-x_0))$; for $x_0=\{0.01,0.1,0.5\}$, $\langle t^*\rangle\approx\{0.1N,0.7N,1.4N\}$, which is consistent with the results in Fig.\ref{Fig:NeutralGreensFunc}. The solution is more simple and intuitive than that of Voronka et al, \cite{Voronka1975}, where it is clear the behaviour of gene frequencies is essentially that of Brownian motion in an unstable harmonic potential; the non-linearity in $x-$space arises purely from working in the more natural co-ordinates of the angular transformation, where in particular the argument of the exponential in the Gaussian solution is just the square of the stochastic distance between $x_0$ and $x$. For example, it is instructive that, as a consequence, even for a neutral process the mean of allele frequencies moves towards the extinction or fixation boundary for any initial frequency $x_0\ne0.5$, as is seen clearly from the plots of the Green's functions in Fig.\ref{Fig:NeutralGreensFunc} and from the solution of the mean $\langle\theta(t)\rangle$; this is not obvious from the neutral Wright-Fisher diffusion equation in $x-$space (Eqn.\ref{Eq:ForwardFokkerPlanck} with $s=0$).

For the case of selection, from Eqn.\ref{Eq:ForwardFokkerPlanck} and using Fisher's angular transformation, the stochastic differential equation for $\theta$ is

\begin{equation}\label{Eq:SDE_Theta_selection}
\frac{\ud\theta}{\ud t}=-\frac{1}{2N}\left(\cot(\theta)-Ns\sin(\theta)\right)+\eta(t),
\end{equation}
where $\eta(t)$ has the same moments as before. Note that the contribution of selection to the effective force tends to zero as $\theta\rightarrow \{0,\pi\}$, which agrees with the intuition that when an allele is rare, the change in allele frequency is dominated by drift; in particular, for $\theta\ll 1$, and $Ns\gg1$, $2Nf(\theta)\approx -1/\theta + Ns\theta$ and the forces of drift and selection are roughly in balance when $Ns\sim1/\theta^2= 1/4x$, where Fisher's angular transformation is $\theta\approx\sqrt{4x}$ for small $x$ -- in other words when the allele frequency $x\ll (4Ns)^{-1}$ drift dominates.

When selection is weak ($Ns\ll1$), the effective force in the angular domain is only a weak perturbation on the neutral force (Fig.\ref{Fig:Neutral+SelectiveForce}) and the Green's functions differ little from neutrality, particularly at short times (not shown). A similar linear expansion of the force can be carried out to calculate an asymptotic expression for the Green's function under weak selection, as shown in the Supplementary Online Material; the resulting expression has similar accuracy compared to numerical simulations as in the neutral case.

In the regime of strong selection $Ns\gtrsim 1$, the above approach gives a poor approximation, due to the non-linearity of the effective force in the angular domain (Fig.\ref{Fig:Neutral+SelectiveForce}). Here we present a heuristic approach to solving Eq.\ref{Eq:SDE_Theta_selection} approximately, for any value of $Ns$. The approach is to assume that the Green's function of the non-linear SDE can be approximated by a Gaussian process, where:  1) the time-varying mean is calculated as a solution to the deterministic dynamics of the SDE Eq.\ref{Eq:SDE_Theta_selection}, with initial condition $\theta_0$, which we show below can be calculated exactly; and 2) the time-varying variance $\langle\langle\theta^2(t)\rangle\rangle$ is dependent on the local gradient of the force, which varies as a function of the deterministic solution, $\lambda=f'(\langle\theta(t)\rangle)$. In general, if an exact solution is not available to the deterministic dynamics, an approximate solution that makes a linear approximation of the effective force (Fig.\ref{Fig:Neutral+SelectiveForce}) about the initial condition, also gives accurate results at short times (not shown).

%and because solutions are relatively insensitive to the mean. On the other hand the Greens function solutions are quite sensitive to the variance as this controls the width of the distribution over time.
Transforming the deterministic part of Eqn.\ref{Eq:SDE_Theta_selection} back to $x-$space, we have $\dot{x}=-\frac{1}{2N}(\frac{1}{2}(1-2x)-2Nsx(1-x)$, the solution to which is of the form $x=C+A\tanh{(\gamma t/2+B)}$. Transforming back to $\theta-$space and using the initial condition $\theta_0=\langle\theta(0)\rangle$, the solution to the deterministic dynamics of Eqn.\ref{Eq:SDE_Theta_selection} is:

\begin{equation}\label{Eq:DeterministicSolutionStrongSelection}
\langle\theta(t)\rangle = \cos^{-1}\left(-\frac{1+2N\gamma\tanh\left(\gamma t/2 -\alpha \right)}{2Ns}\right)
\end{equation}
where $\alpha=\tanh^{-1}\left(\frac{2Ns\cos{\theta_0}+1}{2N\gamma}\right)$ and the characteristic rate of change of the mean is $\gamma=\sqrt{1 + 4N^2s^2}/2N$.

%The constant $\vartheta$ is determined by the initial condition $\langle\theta(0)\rangle=\theta_0$, $\vartheta= (2Ns\cos(\theta_0)+1+\alpha)/(2Ns\cos(\theta_0)+1-\alpha)$.

The next step is to calculate the variance, which we motivate by considering the situation when the slope of the effective force is fixed to a constant $\lambda$, which gives a Gaussian solution with variance $\langle\langle\theta^2(t)\rangle\rangle=\frac{1}{2N\lambda}(e^{2\lambda t}-1)$. The linearity of the force characterises the Gaussian distribution and so if we assume that the effective deterministic force varies slowly over a range of theta representing the width of the probability density, we can then heuristically replace $\lambda$ with the local derivative of the effective force $\lambda(\langle\theta(t)\rangle)$ in the variance. This approximates the local spreading of the probability density being solely due to the local derivative of the force giving a time varying variance:
\begin{equation}\label{Eq:VarThetaSolution_Heuristic2}
\langle\langle\theta^2(t)\rangle\rangle=\frac{1}{2N \lambda(\langle\theta(t)\rangle)}(e^{2 \lambda(\langle\theta(t)\rangle)t}-1).
\end{equation}
Note that for strong selection, the derivative of the effective deterministic force $\lambda(\langle\theta(t)\rangle)$ will be zero at certain times, as can be seen from the plot of the deterministic force in Fig.\ref{Fig:Neutral+SelectiveForce}; at these time points it is simple to see that the variance remains well behaved as $\lim_{\lambda\rightarrow 0}\{\langle\langle\theta^2\rangle\rangle\}\rightarrow t/N$, as one would expect if the deterministic force tends to a constant. Transforming back to $x-$space, the Green's function solution is:
%As long as there are a sufficient number of generations on the timescale the mean changes, i.e. $1/\gamma\gg1$, which for $Ns\gg1$, requires $s\ll1$ (from the definition of $\gamma$), then $\theta(t)$ will be roughly Gaussian with mean and variance given above.

%the timescale that the mean changes and
%\begin{widetext}
\begin{align}\label{Eq:HeuristicGreensSoln}
G_x(x,x_0;t)=\frac{\exp{-\frac{\left(\cos^{-1}(1-2x)-\langle\theta\rangle\right)^2}{2\langle\langle\theta^2\rangle\rangle}}}{\sqrt{2\pi x(1-x)\langle\langle\theta^2\rangle\rangle}},
\end{align}
%\end{widetext}
where $\langle\theta\rangle$ and $\langle\langle\theta^2\rangle\rangle$ are given by Eqns.\ref{Eq:DeterministicSolutionStrongSelection} and \ref{Eq:VarThetaSolution_Heuristic2}, respectively, where $\cos(\theta_0)=1-2x_0$.

%\begin{figure}[htb!]
%\begin{center}
%{\rotatebox{0}{{\includegraphics[width=0.5\textwidth]{Figures/IntSelection_Small_x0.eps}}}}
%\caption{Comparison of approximate calculation of Greens function (solid lines -- Eqn.\ref{Eq:GreensFunc_x_Harmonic_Soln}) and numerical integration of stochastic differential equation that arises from diffusion approximation (solid circles). a) initial frequency $x_0=0.01$, b) $x_0=0.99$.\label{Fig:IntSelectionSmallx0GreensFunc}}
%\end{center}
%\end{figure}

We plot the results for $Ns=10$ ($x_0=\{0.1,0.5,0.9\}$ in Fig.\ref{Fig:StrongSelectionGreensFunc}; see Supplementary Online Information for plots with $x_0=\{0.01,0.99\}$) (Green's functions for $Ns=1$ are plotted in the Supplementary Online Information). We find that for both $Ns=1$ and $Ns=10$ the heuristic approach and the integration of the Wright-Fisher SDE (Eqn.\ref{Eq:SDE_Theta_selection}) agree very well at short times compared to the average expected time for fixation/extinction of a mutant. This is true even when $x_0$ is very close to $0$ or $1$ (Supplementary Online Information) and is reasonably accurate to quite long times ($t\sim\tau$) for an initial frequency of $x_0=0.1$ (Fig.\ref{Fig:StrongSelectionGreensFunc}A). In addition, we see that as well as capturing the broad behaviour of the time-varying mean and variance, the insets of the figures show the Green's functions plotted on a log scale, demonstrating that the approximation is also very accurate in the tails of the distribution at short times. Finally, for very long times when $\langle\theta(t)\rangle$ tends to $0$ or $\pi$, the variance of the heuristic solution Eqn.\ref{Eq:VarThetaSolution_Heuristic2} diverges, as $\lambda$ diverges at the boundaries, and the approximation fails; this is indicated in those cases where there is no heuristic solution plotted for a given time in each plot.

%\begin{figure}[htb!]
%\begin{center}
%{\rotatebox{0}{{\includegraphics[width=0.5\textwidth]{Figures/IntSelection.eps}}}}
%\caption{Comparison of approximate calculation of Greens function (solid lines -- Eqn.\ref{Eq:GreensFunc_x_Harmonic_Soln}) and numerical integration of stochastic differential equation that arises from diffusion approximation (solid circles). a) initial frequency $x_0=0.1$, b) $x_0=0.5$, c) $x_0=0.9$.\label{Fig:IntSelectionGreensFunc}}
%\end{center}
%\end{figure}

\begin{figure}[htb!]
\begin{center}
{\rotatebox{0}{{\includegraphics[width=0.5\textwidth]{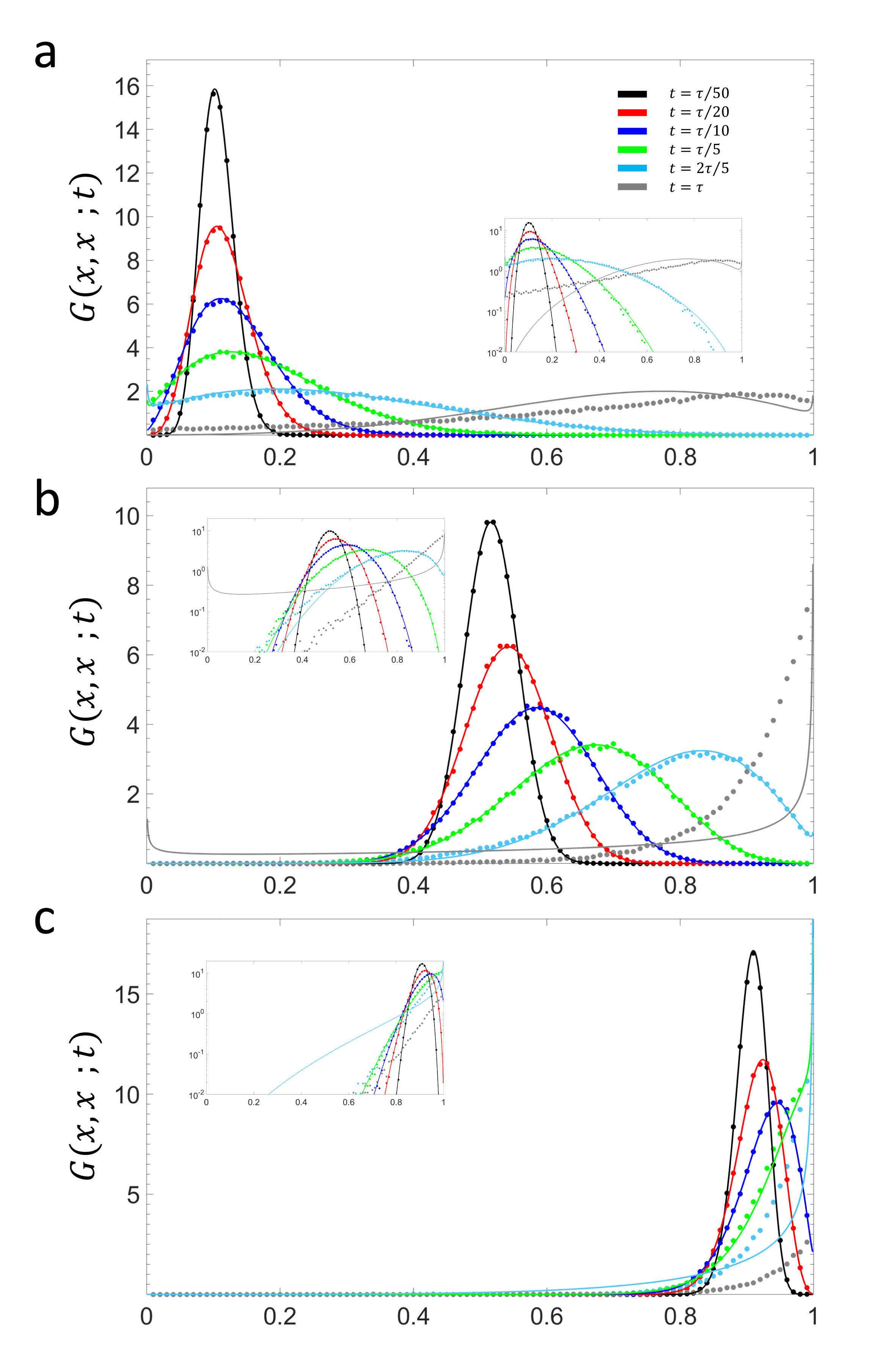}}}}
\caption{Comparison of approximate calculation of Greens function for $Ns=10$ (solid lines -- Eqn.\ref{Eq:HeuristicGreensSoln}) and numerical integration of stochastic differential equation that arises from diffusion approximation (solid circles). a) initial frequency $x_0=0.1$, b) $x_0=0.5$, c) $x_0=0.9$. Green's functions are plotted at times given by fractions of $\tau=\frac{1}{s}(1+\ln{(Ns)})$, which is approximately the expected time to fixation of a mutant which survives drift and then is driven to fixation by selection \cite{Desai2007}. \label{Fig:StrongSelectionGreensFunc}}
\end{center}
\end{figure}

%\begin{figure}[htb!]
%\begin{center}
%{\rotatebox{0}{{\includegraphics[width=0.5\textwidth]{Figures/StrongSelection_Small_x0.eps}}}}
%\caption{Comparison of approximate calculation of Greens function (solid lines -- Eqn.\ref{Eq:GreensFunc_x_Harmonic_Soln}) and numerical integration of stochastic differential equation that arises from diffusion approximation (solid circles). a) initial frequency $x_0=0.01$, b) $x_0=0.99$. Green's functions are plotted at times given by fractions of $\tau=\frac{1}{s}(1+\ln{(Ns)})$, which is approximately the expected time to fixation of a mutant which survives drift and then is driven to fixation by selection \cite{Desai2007}.\label{Fig:StrongSelectionSmallx0GreensFunc}}
%\end{center}
%\end{figure}

%For $Ns=10$, we plot the Green's functions at times given by fractions of $\tau=\frac{1}{s}(1+\ln{(Ns)})$, which is approximately the expected time to fixation of a mutant which survives drift and then is driven to fixation by selection \cite{Desai2007}.

%\section{Conclusions}

To conclude, we have calculated very accurate approximations of the 2-allele Green's function (or transition probability density) of population genetics for arbitrary selection coefficient $s$ and population size $N$. A key advantage and insight of the approach outlined in this paper, is that it transforms a non-linear Fokker-Planck equation to a simple problem of Brownian motion in an effective potential. Together with the heuristic Gaussian approximation, this represents, to the author's knowledge, a new general approach for asymptotically solving Fokker-Planck equation's with a co-ordinate dependent diffusion constant in slowly-varying potentials (or equivalently SDEs with multiplicative noise), where the solution to the mean behaviour is known; indeed, in 1-dimension a PDE with co-ordinate dependent diffusion can always be transformed to one with co-ordinate independent diffusion \cite{Antonelli1977,Baxter2007}. For more than two variants the methods detailed in \cite{Antonelli1977}, suggests via the metric tensor, a potential route to finding solutions in higher dimensions.

These results have potential application to detecting selection in time-series data of the composition of variants, in biological evolution, language evolution and for species in ecosystems. In particular, as these results have accuracy in the asymptotic short-time limit, they will be applicable to studying selection from time-series of variants (haplotypes) in virus evolution, since they have large effective population sizes and short generation times, meaning even sampling virus populations infrequently (on the time scale of many months or years) would be accurately modelled by the results of this paper.

\begin{acknowledgements}
I thank Richard A. Goldstein for initially suggesting the problem and for useful discussions. I also thank Richard Blythe for useful comments on the manuscript. This work was supported by The Francis Crick Institute which receives its core funding from Cancer Research UK, the UK Medical Research Council and the Wellcome Trust.
\end{acknowledgements}

%%GATHER{D:/Documents/LaTeX/BiBTeX/Evolutionpapers.bib}
\bibliography{D:/Documents/LaTeX/BiBTeX/Evolutionpapers}% Produces the bibliography via BibTeX.
%%%%%%%%%% Merge with supplemental materials %%%%%%%%%%
\pagebreak
\widetext
\begin{center}
\textbf{\large Supplemental Materials}
\end{center}
%%%%%%%%%% Merge with supplemental materials %%%%%%%%%%
%%%%%%%%%% Prefix a "S" to all equations, figures, tables and reset the counter %%%%%%%%%%
\setcounter{equation}{0}
\setcounter{figure}{0}
\setcounter{table}{0}
\setcounter{page}{1}
\makeatletter
\renewcommand{\theequation}{S\arabic{equation}}
\renewcommand{\thefigure}{S\arabic{figure}}
\renewcommand{\bibnumfmt}[1]{[S#1]}
\renewcommand{\citenumfont}[1]{S#1}
%%%%%%%%%% Prefix a "S" to all equations, figures, tables and reset the counter %%%%%%%%%%

\section*{Green's function under weak selection}

For the case of selection, using Eqn.1 in the main text Fisher's angular transformation $\theta=\cos^{-1}(1-2x)$, the stochastic differential equation for $\theta$ is

\begin{equation}\label{Eq:SDE_Theta_selection2}
\frac{\ud\theta}{\ud t}=-\frac{1}{2N}\left(\cot(\theta)-Ns\sin(\theta)\right)+\eta(t),
\end{equation}
where $\langle\eta(t)\rangle=0$ and $\langle\eta(t)\eta(t')\rangle=\delta(t-t')/N$. We first try to solve this by expanding the effective deterministic force,
\begin{equation}\label{Eq:DriftForce_Selection}
f(\theta)=\frac{1}{2N}(\cot(\theta)-Ns\sin(\theta)),
\end{equation}
about its zero $\theta^\ast$ and solve the resulting linear SDE as for the neutral case. This only works well for the case of weak selection, as the effective deterministic force becomes increasingly non-linear for all values of $\theta$ when selection is strong (Fig.1 main text). The solution to $f(\theta)=0$ is

\begin{equation}\label{Eq:Selection_ZeroForce}
\theta^\ast=\cos^{-1}\left(\frac{-1+\sqrt{1+4N^2s^2}}{2Ns}\right),
\end{equation}
which for weak selection, $4Ns\ll1$, is simply,
\begin{equation}\label{Eq:WeakSelection_ZeroForce}
\theta^\ast\approx\cos^{-1}\left(Ns\right).
\end{equation}
Intuitively, this makes sense, as for $s>0$, this gives $\theta^\ast<\pi/2$, so the dividing point (separatrix) between initial conditions that result in deterministic dynamics giving fixation of the mutant allele is shifted to smaller values of $\theta_0$ compared to neutral ($\theta_0=\pi/2$); the converse is true, for $s<0$, where $\theta^\ast>\pi/2$. The force expanded to linear order is $f(\theta)=\lambda(\theta-\theta^\ast)$, where
\begin{equation}\label{Eq:Selection_Force_deriv}
\lambda=f'(\theta^*)=\frac{1}{2N}\left(\frac{1}{1-\cos^2(\theta^*)}+Ns\cos(\theta^*)\right),
\end{equation}
which again for weak selection is approximately
\begin{equation}\label{Eq:Weak_Selection_Force_deriv}
\lambda=f'(\theta^\ast)\approx\frac{1}{2N}\left(1+2N^2s^2\right).
\end{equation}
The stochastic equation of motion is then
\begin{equation}\label{Eq:SDE_Theta_Harmonic_WeakSelection}
\frac{\ud\theta}{\ud t}=\lambda\left(\theta-\theta^\ast\right)+\eta(t).
\end{equation}
The analysis then proceeds in the same way as for the neutral case giving the Green's function in $x-$space as
\begin{widetext}
\begin{align}\label{Eq:GreensFunc_Harmonic_WeakSelection}
G_x(x,x_0;t)=\sqrt{\frac{N\lambda}{\pi x(1-x)(e^{2\lambda t}-1)}}\exp{-\frac{\left(\cos^{-1}(1-2x)-\cos^{-1}(1-2x_0)e^{\lambda t}-\theta^\ast(1-e^{\lambda t})\right)^2}{\left(e^{2\lambda t}-1\right)/N\lambda}},
\end{align}
\end{widetext}
In the regime where $Ns\ll1$, stochastic simulations and Eqn.\ref{Eq:GreensFunc_Harmonic_WeakSelection} agree well at short times with a similar accuracy (not shown) as shown in Fig.2 in the main text; the results show the Green's function under weak selection are only a small perturbation on the neutral Green's function Eqn.6 in main text, and only diverge significantly for long times where this approximation, in any case fails.

\section*{Supplementary Figures}

\begin{figure}[h]
\begin{center}
{\rotatebox{0}{{\includegraphics[width=0.5\textwidth]{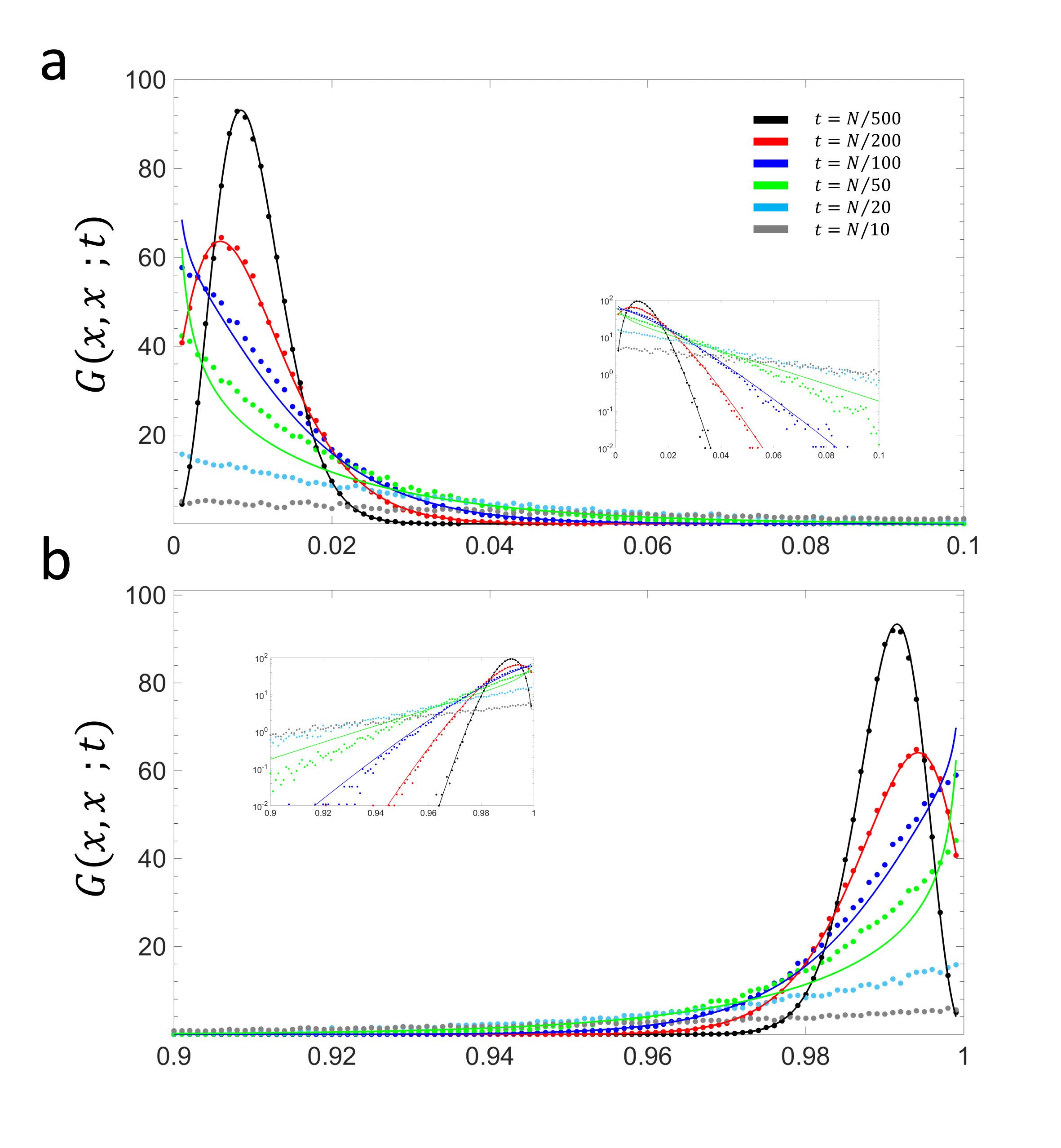}}}}
\caption{Comparison of approximate calculation of Greens function for $Ns=1$ (solid lines -- Eqn.10 in main text) and numerical integration of stochastic differential equation that arises from diffusion approximation (solid circles). a) initial frequency $x_0=0.01$, b) $x_0=0.99$.\label{Fig:IntSelectionSmallx0GreensFunc}}
\end{center}
\end{figure}

\begin{figure}[h]
\begin{center}
{\rotatebox{0}{{\includegraphics[width=0.5\textwidth]{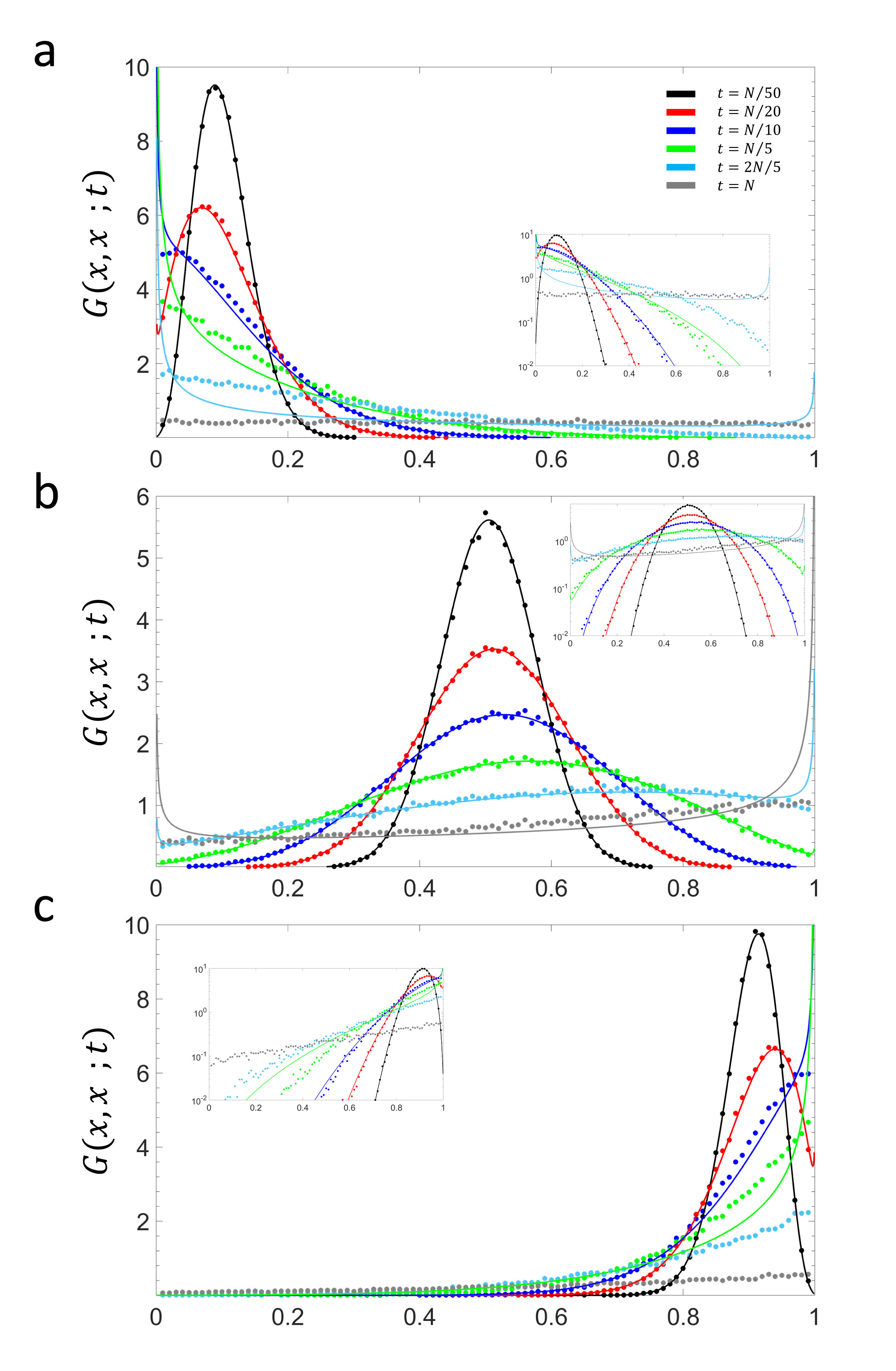}}}}
\caption{Comparison of approximate calculation of Greens function for $Ns=1$ (solid lines -- Eqn.10 in main text) and numerical integration of stochastic differential equation that arises from diffusion approximation (solid circles). a) initial frequency $x_0=0.1$, b) $x_0=0.5$, c) $x_0=0.9$.\label{Fig:IntSelectionGreensFunc}}
\end{center}
\end{figure}

We plot the results for $Ns=1$ ($x_0=\{0.01,0.99\}$ in Fig.\ref{Fig:IntSelectionSmallx0GreensFunc} and $x_0=\{0.1,0.5,0.9\}$ in Fig.\ref{Fig:IntSelectionGreensFunc}). As discussed in the main text, we find that for both $Ns=1$ and $Ns=10$ the heuristic approach and the integration of the Wright-Fisher SDE (Eqn.7 in main text) agree very well at short times compared to the average expected time for fixation/extinction of a mutant.

For $Ns=1$, we expect the mean time to fixation/extinction to be of order $\sim N=1/s$ and so the Green's functions are plotted at different times $t$ which are fractions of $N$. We see that the time range from zero that the approximation is accurate decreases as the initial frequency $x_0$ is nearer to either of the boundaries, but for sufficiently short times, even when $x_0=0.01$ or $x_0=0.99$ (Fig.\ref{Fig:IntSelectionSmallx0GreensFunc}), the heuristic solution is very accurate. The main difference in the Green's functions for $Ns=1$ and $Ns=10$ are that the distributions are more narrow about their peak for $Ns=10$, which is as expected as under stronger selection as the dynamics will be more deterministic; we see that the heuristic approximation captures this behaviour very accurately.

\begin{figure}[h]
\begin{center}
{\rotatebox{0}{{\includegraphics[width=0.5\textwidth]{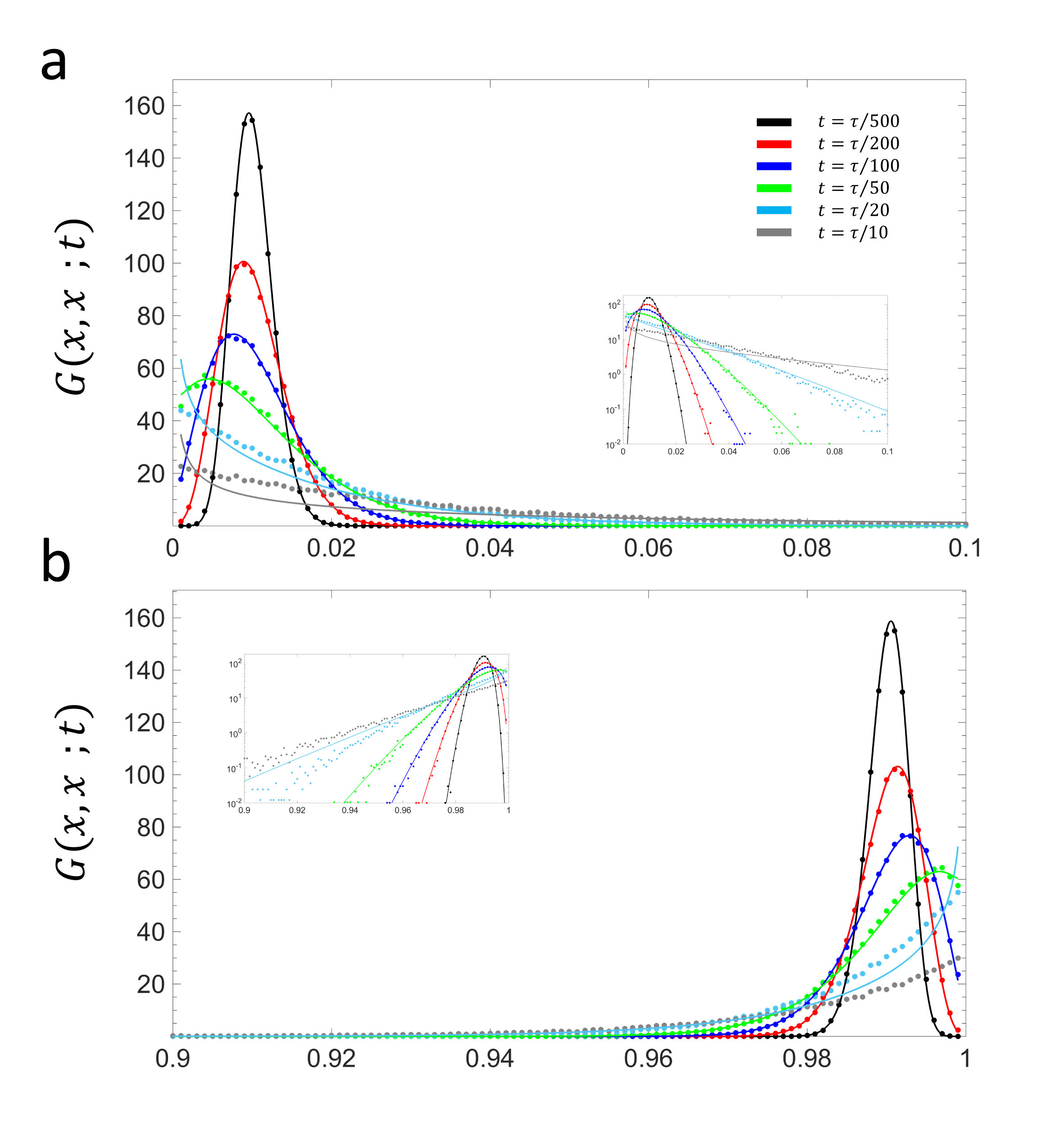}}}}
\caption{Comparison of approximate calculation of Greens function for $Ns=10$ (solid lines -- Eqn.10 in main text) and numerical integration of stochastic differential equation that arises from diffusion approximation (solid circles). a) initial frequency $x_0=0.01$, b) $x_0=0.99$. Green's functions are plotted at times given by fractions of $\tau=\frac{1}{s}(1+\ln{(Ns)})$, which is approximately the expected time to fixation of a mutant which survives drift and then is driven to fixation by selection.\label{Fig:StrongSelectionSmallx0GreensFunc}}
\end{center}
\end{figure}

In Fig.\ref{Fig:StrongSelectionSmallx0GreensFunc}, we have plotted the approximate heuristic Green's function for strong selection ($Ns=10$), which initial frequencies of $x_0=\{0.01,0.99\}$. We see that the approximation is again very good for sufficiently short times compared to the expect time to fixation or extinction.

\end{document}